**An Evaluation of Bounding Approaches for Generalization**

Wendy Chan

Human Development and Quantitative Methods Division

Graduate School of Education, University of Pennsylvania, Philadelphia, PA

Email: wechan@upenn.edu

ORCiD: https://orcid.org/0000-0002-0933-9532

**An Evaluation of Bounding Approaches for Generalization**

**Abstract**

Statisticians have recently developed propensity score methods to improve generalizations from randomized experiments that do not employ random sampling. However, these methods typically rely on assumptions whose plausibility may be questionable in practice. In this article, we introduce and discuss bounding, an approach that is based on alternative assumptions that may be more plausible in a given study. The bounding framework nonparametrically estimates population parameters using a range of plausible values that are consistent with the observed characteristics of the data. We illustrate how the bounds can be tightened using three approaches: imposing an alternative assumption based on monotonicity, redefining the population of inference, and using propensity score stratification. Using the results from two simulation studies, we examine the conditions under which bounds for the population parameter are tightened. We conclude with an application of bounding to SimCalc, a cluster randomized trial that evaluated the effectiveness of a technology aid on mathematics achievement.

**Introduction**

The past decade has seen a rise in the use of experiments to evaluate the causal impact of interventions in fields such as education, psychology, economics and sociology. Although experiments ensure an important aspect of internal validity through treatment randomization, many social experiments lack external validity, or generalizability, when the study samples are not a random subset of the population of interest (Shadish, Cook, & Campbell, 2002). Probability, or random sampling, is the strongest tool for statistical generalization, but this sampling is often not done in practice for reasons that include limited resources and the fact that the populations of interest are not defined a priori (O'Muircheartaigh and Hedges, 2014). In a review of experiments listed in the *Digest of Social Experiments*, Olsen, Orr, Bell and Stuart (2013) found that less than 3% of the studies used both random treatment assignment and random sampling (Greenberg and Shroder, 2004). Without random sampling, estimates of the average treatment effect for the population are no longer unbiased, which can lead to potentially misleading findings. Such outcomes have important implications for policy.

      Statisticians recently developed methods to improve generalizations using propensity scores. Propensity scores match subjects in the experimental study with subjects in the population so that any difference between the two groups is not systematic (Rosenbaum and Rubin, 1983). Although propensity score methods have made an important contribution to causal generalization, they require several assumptions for their validity. In particular, propensity scores for generalization require *sampling ignorability*, which assumes that (1) the propensity scores contain all possible covariates that explain treatment effect variation and affect sample selection so that any remaining differences between the study sample and population can be ignored; and (2) every subject in the sample has a "comparable" subject in the population, where comparability is assessed using characteristics that are observable for each subject in the sample

and the population. Whether sampling ignorability is plausible in any given study is often questionable because the assumption relies on conjectures on the relationship between sample selection and the expected treatment effect. Importantly, situations in which sampling ignorability fails can occur, and this has implications for statistical inference.

In practice, researchers have several options to address violations in sampling ignorability. Two approaches, which we discuss briefly in this paper, are sensitivity analyses and, specifically for our context, redefining the population for generalization to improve the plausibility of sampling ignorability. The purpose of this article is to introduce and discuss a third option, bounding, in which the population parameter is nonparametrically estimated by a range of plausible values that are consistent with the observed characteristics of the data. Bounding approaches are motivated by exploring the types of inferences under alternative assumptions that are potentially more plausible in a study (Manski, 2009). The goal of this article is two-fold. First, we introduce the bounding framework for generalization and discuss the advantages and limitations of this framework for studies where the assumption of sampling ignorability may be violated. Second, because an important limitation of bounding methods is that the estimated bounds are often uninformatively wide, we describe three methods to improve the precision of bounds: imposing a monotonicity assumption, redefining the population of inference, and propensity score stratification. Throughout this analysis, we identify the conditions under which each method yields precision gains in the estimated bounds.

The article is organized as follows. We first introduce SimCalc, a cluster randomized trial on a mathematics technology aid, that serves as the motivating example with which we base our discussion on causal generalization. We then introduce the notation and assumptions for propensity scores and discuss the plausibility of sampling ignorability in SimCalc. We introduce bounding and illustrate how bounds can be tightened using two simulation studies and through a re-analysis of our motivating example. Finally, we provide some concluding thoughts about the tradeoffs of bounding as well as ideas for future research.

**SimCalc**

SimCalc is a mathematics software program that uses computer animations to teach concepts of rate and proportions. A mission of the SimCalc Project, based at the James J. Kaput Center at the University of Massachusetts, Dartmouth, is to provide students in disadvantaged environments with opportunities to learn advanced mathematics (Kaput, 1997). In addition to the software, the SimCalc Project also provided professional development workshops for teachers to strengthen their mathematical content knowledge, to learn to use the curriculum materials associated with SimCalc, and to specifically plan for the use of the materials (Roschelle et al., 2010, p. 847). To assess the impact of SimCalc on mathematics achievement, the research firm SRI International implemented two cluster randomized experiments, one of which was a pilot study, on a combined sample of 92 middle schools in Texas. In the original study, the principal investigators found a statistically significant main effect of 1.438 (SE = 0.143, $p < 0.001$) in students' gain scores, implying that students in schools using SimCalc experienced larger gains than students in control schools (Roschelle et al., 2010). This effect is standardized in relation to the between-school variance.

Although every effort was made to select a random sample of schools, the SimCalc sample was not a probability sample of Texas middle schools. Schools were primarily recruited through the Charles A. Dana Center at the University of Texas and regional education service centers throughout Texas (Roschelle et al., 2010, p.855). The generalization question of interest is, if SimCalc were implemented statewide in Texas, what is the expected impact for a "typical" school? Because the 92 schools that participated in the study may not necessarily be representative of all Texas middle schools, generalizing the results of SimCalc requires estimation methods that rely on several assumptions.

Tipton (2013) and O'Muircheartaigh and Hedges (2014) both addressed this generalization question using a subclassification estimator based on propensity scores. In both studies, propensity scores were used to match the SimCalc and Texas population schools on 26 covariates deemed relevant in predicting sample selection and the treatment effect. These covariates were taken from the Texas Academic Excellence Indicator System (AEIS) and summary statistics are given in Table 1. Tipton (2013) estimated an average treatment effect of 1.452 (SE = 0.195, $p < 0.01$) in the original population and an effect of 1.430 (SE = 0.188, $p < 0.01$) in a subpopulation of schools that had the greatest similarity in covariate distributions to the SimCalc sample. Both estimates imply that a typical population school using SimCalc would experience an overall average gain in students' scores.

Several assumptions were needed to derive the point estimates for SimCalc and an important question is whether the assumptions are plausible and the implications for inference if these core assumptions do not necessarily hold. In the next section, we introduce the framework and notation used throughout the article. We discuss the assumptions needed to use propensity scores for causal generalization and describe conditions under which the assumptions would not hold in practice.

TABLE 1

**Notation and Assumptions**

To describe the identification problem of interest, we define the treatment effect based on a finite population potential outcomes framework (Imbens and Rubin, 2015). Consider a population $P$ of $N$ schools of which a sample $n$ is selected into the experimental study. In SimCalc, $P$ consists of 1,713 non-charter schools that serve seventh grade students, of which $n = 92$ schools were selected into the study. Following the empirical example, we assume throughout that the sample of $n$ schools was not randomly selected from $P$, but that within the study, the schools were randomly assigned to each treatment condition. For each school $i$, $i = 1, ..., N$, let $W_i$ be a binary treatment assignment indicator with $W_i = 1$ if school $i$ was randomized to the treatment condition and $W_i = 0$ otherwise. Let $Z_i$ indicate selection into the experimental sample where $Z_i = 1$ if school $i$ was selected into the sample and $Z_i = 0$ otherwise. Finally, let $Y(1), Y(0)$ denote the potential outcomes under the treatment and control conditions, respectively (Rubin, 1977). The treatment effect for school $i$ is the difference in potential outcomes, $\Delta_i = Y_i(1) - Y_i(0)$.

To estimate $\Delta_i$, we assume that the stable unit treatment value assumption (SUTVA) holds for the sample and population (Tipton, 2013; Rubin, 1978, 1980, 1986). Under SUTVA for the sample, SUTVA(S), the potential outcomes for school $i$, $Y_i(1), Y_i(0)$, depend only on the treatment received by $i$ and not on the treatment received by school $j$ for $i \neq j$. SUTVA(S) implies that there are no peer effects in the sense that the response to treatment among students in one school does not depend on the treatment assigned to students in a different school. Additionally, SUTVA(S) stipulates that there is only one version of the treatment. SUTVA for the population, SUTVA(P), is unique to generalization studies and requires that the conditions for treatment assignment under SUTVA(S) hold for sample selection. SUTVA(P) also requires that the potential outcomes do not depend on the proportion of schools selected into the study, and that there is no interference between schools, both among the treatment and control groups and among the sampled and non-sampled groups.

Under SUTVA(S) and SUTVA(P), the sample average treatment effect $\Delta_{\text{SATE}}$ (SATE) is:

$$\Delta_{\text{SATE}} = E(\Delta_i | Z = 1) = E(Y_i(1) - Y_i(0) | Z = 1) \qquad (1)$$

When treatment is randomly assigned, an unbiased estimator of $\Delta_{SATE}$ is given by $\widehat{\Delta}_{SATE} = \frac{1}{n}\sum_{j\in\{Z=1\}} Y_j(1) - Y_j(0)$. In generalization studies, the parameter of interest is the population average treatment effect $\Delta_{PATE}$ (PATE), defined as:

$$\Delta_{PATE} = E(\Delta_i|Z=1)\Pr(Z=1) + E(\Delta_i|Z=0)\Pr(Z=0) \qquad (2)$$

In the SimCalc study, the PATE refers to the expected impact of SimCalc for the population of seventh graders in the given academic year. From Equations (1) and (2), note that the SATE and PATE are equivalent only when treatment effect heterogeneity and sample selection are independent (Imai et al., 2008; Rubin, 1974). This is satisfied under probability sampling when $E(\Delta|Z=1) = E(\Delta|Z=0)$ or when the treatment effects are constant. This condition also holds if $\Pr(Z=1) = 1$, where every school is selected into the experimental study. Otherwise, an estimate $\widehat{\Delta}_{SATE}$ may be unbiased for the SATE, but not necessarily for the PATE.

*Propensity Scores*

Propensity scores model the probability of sample selection as a function of observable characteristics or covariates of schools in both the sample and the population. For each school $i$ in $P$, let $\mathbf{X}$ be a vector of observable covariates, which may include a combination of continuous and categorical variables. The sampling propensity score is defined as $s(X) = \Pr(Z=1|X)$, the conditional probability of selection into the sample. Figure 1 illustrates the distribution of propensity scores in the SimCalc study. Note that the figure shows the *non-sampling propensity scores* $\Pr(Z=0|X)$, where schools in the right tail of the distribution have the smallest probability of selecting into the SimCalc study.

FIGURE 1

A common method of estimating $s(X)$ is with logistic regression, but the propensity scores can also be estimated using multilevel data by including covariates at the school and district level

(Arpino & Mealli, 2011). Once estimated, the propensity scores $s(X)$ are used to match schools in the sample and population so that the resulting groups are compositionally similar.

Several assumptions are needed to use propensity scores for generalization. First, SUTVA(S) and SUTVA(P) must hold for the sample and population. Second, treatment assignment must be strongly ignorable given the propensity scores (Stuart et al., 2011):

$$Y(1), Y(0) \perp W | Z = 1, s(X) \text{ and } 0 < \Pr(W = 1 | Z = 1, s(X)) < 1 \qquad (3)$$

Under this assumption, the potential outcomes are conditionally independent of treatment assignment and every school has a non-zero probability of being assigned to the treatment condition. Third, unconfounded sample selection must hold where the treatment effect is conditionally independent of sample selection:

$$\Delta = Y(1) - Y(0) \perp Z | s(X) \qquad (4)$$

This assumption also requires that **X** contains all covariates that moderate treatment effect variation and sample selection. Finally, sampling ignorability must hold:

$$\Delta = [Y(1) - Y(0)] \perp Z | s(X) \text{ and } 0 < s(X) \leq 1 \qquad (5)$$

This assumption requires that unconfounded sample selection is met and that the distribution of covariates **X** among schools in the sample and the population share common support (Tipton, 2014). The latter condition is satisfied when every school in the population (sample) has a relevant comparison school in the sample (population), and no school has $s(X) \approx 0$.

*Violations of Sampling Ignorability*

One of the goals of this article is to highlight the implications for inference when core assumptions may be violated and to explore the advantages and limitations of alternative estimation frameworks. Of the four assumptions, sampling ignorability is arguably the most challenging to meet in practice and it can be violated for several reasons. First, if any covariates that moderate treatment effects and sample selection are omitted, sampling ignorability does not hold. For example, if larger and more urban schools responded differently to SimCalc compared to smaller and less urban schools, sampling ignorability would not hold if either school size or urbanicity, or both, were omitted from the propensity score model. Sampling ignorability would also fail if some schools in the sample (population) have no comparable schools in the population (sample). For example, if the SimCalc study consisted of all rural schools, but the population contained schools of differing urbanicities, sampling ignorability would not hold as some sample schools may not have a relevant comparison school in the population. The concern is that the sample and population may differ on important characteristics that, when not accounted for or when unobserved, yield estimates of $\Delta_{PATE}$ that can be seriously biased.

From Figure 1, the distribution of propensity scores in the population is notably skewed compared to the distribution in the sample, suggesting that some schools in the population of Texas middle schools may not be well matched with the SimCalc sample. The non-overlap in propensity scores implies a lack of common support, which limits the extent of generalization (Dehejia & Wahba, 1999, 2002). In practice, non-overlap in propensity score distributions may be attributed to specific covariates. For example, if urbanicity was strongly predictive of the estimated propensity scores and the distribution of this variable differed widely between the sample and population, non-overlap would occur in the propensity score distributions. Non-overlap may also be driven by the collective differences among several covariates. Importantly, non-overlap suggests that the results of SimCalc may generalize well to some schools in the population but not others.

**Addressing Violations to Sampling Ignorability**

*Sensitivity Analysis*

One approach to address violations in sampling ignorability is to conduct a sensitivity analysis. In observational studies, sensitivity analyses quantify the impact of unmeasured confounding on estimates of the treatment impact (Cornfield et al., 1959). Rosenbaum (2002, 2010) presented three approaches that identify the cutoffs in the relationship between the unobserved confounder and outcome that would change the statistical significance of the estimated treatment effect. VanderWeele and Arah (2011) presented an approach to obtain estimates of the "true" outcome-treatment association and 95% confidence intervals. Nguyen et al. (2018) proposed several sensitivity analysis methods for generalization studies when a covariate that moderates the treatment effect is observed in the sample, but not in the population.

While sensitivity analyses are useful in assessing the impact of violations in assumptions, they involve their own assumptions. For example, many approaches assume that the relationships between the unobserved confounders, outcome, and treatment do not vary given the observed covariates and that there is no three-way interaction among the three variables (Harding, 2003). These assumptions are not invoked in all sensitivity analyses, but they are important in deriving simpler expressions of the bias induced by unmeasured confounding.

*Redefining the Population*

Another approach to address violations in sampling ignorability is to redefine the population. This method was used for SimCalc in Tipton (2013) where population schools whose propensity scores exceeded the maximum propensity score in the SimCalc sample were excluded. Redefining the population is designed to improve the plausibility of sampling ignorability and to potentially improve generalizations if the covariates used to redefine the populations also moderate treatment effects and are predictive of sample selection. However, it is important to note that this approach is based on the *observed* covariates in the sample and population.

**Bounding**

As a third approach, this article focuses on bounding methods for generalization when sampling ignorability does not hold. Bounding methods were first discussed in Manski (1990) who recommended that the statistical analysis should start with the data alone so that researchers approach the problem from the same starting point. Bounding methods typically make few to no assumptions on the data, allowing the researcher to derive nonparametric estimates of the range of values. However, these bounds can be uninformatively wide. In the following sections, we introduce bounding and illustrate how the bounds can be tightened using three methods: (1) invoking a monotonicity assumption, (2) redefining the population of inference and (3) propensity score stratification. Throughout, we identify the conditions under which each approach may lead to tighter bounds.

*Derivation of Bounds*

Consider the simple case where the outcome Y is continuous but bounded in a known range $[Y^L, Y^U]$ in both the sample and the population. This would occur, for example, in studies where the outcome is a test score that is computed using the same scale among all schools. It is assumed that the range $[Y^L, Y^U]$ is the same between the sample and the population. [1]

To derive the bounds for the PATE, we first assume SUTVA and strong ignorability of treatment assignment but exclude sampling ignorability. We also assume that the outcomes are measured without error in the sample and population. Using the law of iterated expectations, the expected values of the potential outcomes in $\Delta_{PATE}$ are given by:

$$E(Y(w)) = \sum_Z \sum_W E(Y(w)|W = w, Z = z) \Pr(W = w, Z = z) \text{ for } w, z = 0,1 \quad (6)$$

---

[1] We argue that this assumption is plausible because the outcome Y is assumed to be quantified and measured in the same way between the sample and the population. In some cases, it is possible that the range $[Y^L, Y^U]$ may be different between the sample and the population. For example, if Y was an assessment score and the population was defined at a different time point, the range may be different if the scale of the assessment changes between time points. Alternatively, the range may also be different if the sample was not a subset of the population. If the sample included all general education students, but the population comprised of students in special education classes, the range $[Y^L, Y^U]$ may also be different. However, our framework does not consider these cases because we believe the former is an example of when the outcome Y is defined differently between the sample and the population and the latter addresses a different type of generalization question of interest. If the range $[Y^L, Y^U]$ does differ between the sample and the population, we suggest using the minimum and maximum of Y in the population to derive tight bounds on the PATE.

The expected values of E(Y(1)), E(Y(0)) are weighted combinations of four quantities, corresponding to the treatment indicator W and the selection indicator Z. Of the four terms, three are unobservable counterfactuals (Dawid, 2000; Greenland et al., 1999). In the causal inference literature, counterfactuals refer to the potential outcomes under an experimental condition that is different from the actual condition to which the school was assigned. In (6), the terms E(Y(1)|W=0, Z=1) and E(Y(0)|W=1, Z=1) are treatment counterfactuals and represent the expected outcome under treatment (control) when assigned control (treatment). The terms E(Y(1)|W=1, Z=0), E(Y(1)|W=0, Z=0), E(Y(0)|W=1, Z=0) and E(Y(0)|W=0, Z=0) are sample counterfactuals and represent the expected outcomes for schools not selected into the experimental study. Both sets of counterfactuals are unobservable, but for different reasons. Treatment counterfactuals are unobservable because each school receives at most one treatment, and so the outcome under the alternative treatment condition is unobserved. Sample counterfactuals are unobservable because schools that did not participate in SimCalc do not receive treatment so their potential outcomes under an experimental condition are unobserved.

If sampling ignorability does not hold, the PATE cannot be point identified, but it can be bounded using known features of the data. Specifically, the PATE can be bounded by replacing the unobservable sample counterfactuals with $Y^L$ and $Y^U$ to derive the "worst-case bounds" on the $\Delta_{\text{PATE}}$. This is equivalent to bounding the expected outcomes of the schools that did not participate in SimCalc with the observable bounds on the gain scores. Under this approach, the worst-case bounds for the PATE are given by:

$$\Delta_{\text{PATE}} \in [\Delta^L, \Delta^U] \qquad (7)$$
$$\Delta^L = \Delta_{\text{SATE}} \Pr(Z=1) + (Y^L - Y^U) \Pr(Z=0)$$
$$\Delta^U = \Delta_{\text{SATE}} \Pr(Z=1) + (Y^U - Y^L) \Pr(Z=0)$$

where $\Delta_{SATE} = E(Y(1)|W = 1, Z = 1) - E(Y(0)|W = 0, Z = 1)$. The bounds in (7) are the tightest bounds based solely on the data and knowledge of the range $[Y^L, Y^U]$. However, the width of the bound is $2(Y^U - Y^L) \Pr(Z = 0)$, which, as Chan (2017) pointed out, is never informative of the sign of the PATE.

*Monotone Sample Selection*

To tighten the bounds in (7), we explore the plausibility of a monotonicity assumption on the sample counterfactuals. Prior studies have considered a similar assumption, monotone treatment selection, where schools select the treatment that yields the better potential outcome (Manski, 1990). As an extension to this idea for generalization, we assume that in the absence of probability sampling, if schools choose to participate in the study (*Z=1*) over not participating (*Z=0*), this implies $\Delta = Y(1) - Y(0)|Z = 1 \geq Y(1) - Y(0)|Z = 0$. In other words, a school that chooses to participate in the study does so with the expectation of realizing a larger treatment impact. We refer to this assumption as *monotone sample selection* (MSS). Formally, we define MSS as:

$$Z = 1 \text{ implies } \Delta = Y(1) - Y(0)|Z = 1 \geq \Delta|Z = 0 \qquad (8)$$

MSS improves upon the bound in (7) by replacing the unobserved difference $Y(1) - Y(0)|Z = 0$ with $Y(1) - Y(0)|Z = 1$ instead of $Y^L, Y^U$ for each unobserved potential outcome. As a result, bounds under MSS are narrower because the upper bound in (7) is replaced with a smaller value. If MSS holds, this leads to the following bounds on the $\Delta_{PATE}$:

$$\begin{aligned}\Delta^L &= \Delta_{SATE} \Pr(Z = 1) + (Y^L - Y^U) \Pr(Z = 0) \\ \Delta^U &= \Delta_{SATE} = E(Y(1)|W = 1, Z = 1) - E(Y(0)|W = 0, Z = 1)\end{aligned} \qquad (9)$$

Note that (7) and (9) have the same lower bound since MSS only applies to the upper bound.

An important question is whether MSS is plausible in a given study. While MSS, like sampling ignorability, is not a testable assumption, we argue that its plausibility may be suggested by the context of the study. For SimCalc, we argue that MSS is plausible for four reasons. First, the study was conducted in partnership with the Charles A. Dana Center at the University of Texas, which had a strong commitment to increase the number of students who progress towards advanced placement mathematics courses (Roschelle et al., 2010). Second, the professional development program in the SimCalc study was offered through the Center and these programs were highly regarded throughout the entire state of Texas. The Center's state-wide reputation for high quality professional development programs, coupled with its strong investment in supporting students' mathematics achievement, potentially motivated teachers and school administrators to participate in SimCalc. Third, schools were partly recruited by regional Education Service Centers (ESCs) that provided support for schools and districts in their respective regions. Because the ESCs developed strong positive relationships with the teachers in their networks, these teachers were likely more willing to participate in ESC-supported studies. Finally, the Center had a history of successfully providing large scale professional development programs throughout the state of Texas so that concerns about the feasibility of the study from school administrators and teachers were likely minimized. Additionally, because of its successful history with professional development programs, the Center had the credibility to address teacher and administrator concerns about participation in the SimCalc study.

These arguments only suggest MSS' plausibility and they can certainly be refuted in contexts where the assumption is clearly violated or if the monotonicity condition assumes the opposite direction (that is, the average treatment impact is *greater* for schools that choose not to participate). For example, MSS would not necessarily hold if schools that chose to participate do so to obtain free resources and not necessarily to realize a larger expected treatment effect. The challenges in validating MSS highlight an important limitation of the bounding framework, but this limitation is not unique to bounding since propensity score methods require their own assumptions. In practice, the plausibility of assumptions, whether it is MSS or sampling ignorability, relies on the researchers' knowledge of the study context and the mechanisms through which schools select to participate in an experimental study.

*Redefining the Population and Propensity Score Stratification*

MSS narrows the bounds in (7), but they may still be wide if the range $Y^L - Y^U$ is wide or the probability $\Pr(Z = 0)$ is large. To tighten the bounds in (7) and (9) further, we explore two additional methods: redefining the population and stratifying the population using propensity scores. As described previously, redefining the population creates a subgroup of schools that shrinks the magnitude of $\Pr(Z = 0)$ in (7) and (9), thereby shrinking the width of the bounds. Although redefining the population limits the extent of generalization, this approach improves the precision of the bounds for a specific subpopulation of inference.

Alternatively, the bounds in (7) and (9) can be tightened by stratifying the population and sample schools using covariates that moderate the treatment effect and sample selection. The goal of stratification is to divide the SimCalc and population schools into subgroups that are compositionally similar in covariate distribution, estimate the bounds in each stratum, and derive a general bound by averaging across the strata. Covariate stratification has been used in prior studies to improve the precision of bounds. Lee (2009) used this approach to tighten bounds on the impact of Job Corps on total earnings. Long and Hudgens (2013) evaluated conditions under which covariates tightened bounds for principal effects. Recently, Miratrix et al. (2018) examined the effect of covariate stratification on bounds for principal causal effects.

In our study, we examine the role of *propensity score* stratification on improving the precision of bounds. Note that propensity scores are not used to derive point estimates in this context, and so we do not assume sampling ignorability. Rather, propensity score stratification is used to create matched subgroups in the population where each subgroup shares similar propensity score distributions and consequently, similar covariate distributions. This similarity is leveraged to create tighter bounds by dividing the schools in the sample and population based on variables that are predictive of sample selection and the outcome.

To derive bounds under stratification, let $P$ be divided into $k$ strata, each of which contains $N_k$ population schools and $n_k$ sample schools. Let $S_j$ define a school's stratum membership where $S_j \in \{1, 2, \ldots, k\}$. Within each stratum $k$, the PATE is defined as $\Delta_{\text{PATE}}^k = \sum_{j:S_j=k} \frac{1}{N_j}(Y_j(1) - Y_j(0))$. The overall bound of the PATE is derived by averaging the lower and upper bounds across the strata:

$$\Delta_{\text{PATE}} \in [\Delta^L, \Delta^U] \tag{10}$$

$$\Delta^L = \sum_{j:S_j=1}^{k} \frac{N_j}{N}\left(E\left(Y_j(1)\right)^L - E\left(Y_j(0)\right)^U\right)$$

$$\Delta^U = \sum_{j:S_j=1}^{k} \frac{N_j}{N}\left(E\left(Y_j(1)\right)^U - E\left(Y_j(0)\right)^L\right)$$

where $N_j/N$ is the proportion of population schools in stratum $j$, $j=1, \ldots, k$, and $E(Y)^L$, $E(Y)^U$ are the lower and upper bounds of the expected potential outcomes, respectively. Stratified versions of the worst-case and MSS bounds can be estimated directly using (7) and (9), and by replacing the whole sample bounds $E(Y)^L, E(Y)^U$ with their stratum-specific equivalents.

When the population is stratified into equally sized strata, the fraction $N_j/N$ is equivalent to $1/k$. Although $P$ can be stratified in different ways, stratification with equally sized strata has been shown to be optimal in terms of bias reduction and ease of implementation (Cochran, 1968). Regarding $k$, both Cochran (1968) and Rosenbaum & Rubin (1984) found that using $k = 5$ strata was sufficient to reduce up to 90% of the bias in covariates. For this reason, our simulation studies are based on *k* = 5 strata. However, it is important to note that five strata are not always possible, particularly when it leads to strata that have no sample schools. In these cases, the number of strata should be reduced to ensure that each stratum contains sample schools.

**Simulations**

We conducted two simulation studies to illustrate the bounding framework. In both studies, the population *P* consists of *N* = *2000* schools of which *n* = *100* are selected into the sample. This places Pr(Z=1) = 100/2000 = 0.05, or 5%, of the population in the experimental sample, which is about the same estimated proportion in SimCalc. To identify the conditions under which bounds for the PATE can be tightened, we varied three main simulation parameters: the correlation among covariates in the propensity score model, the plausibility of MSS, and the definition of the population of inference.

*Correlation Among Covariates*

For both simulation studies, we generated four continuous covariates, $X_1, X_2, X_3, X_4$, from a multivariate normal distribution with mean zero and variance-covariance matrix with main diagonal elements of 1 and the correlations specified as follows: $\rho_{X_1,X_2} = 0.5, \rho_{X_1,X_3} = \rho_{X_2,X_4} = \rho$, where $\rho = 0.25, 0.50, 0.70$ and all other pairwise correlations set to 0.05. We specified the correlation structure in this way because the covariates used in many generalization studies are often at least weakly correlated. Using $X_1, X_2, X_3, X_4$, the selection model is given by:

$$\Pr(Z = 1) = \text{expit}(\beta_1 X_1 + \beta_2 X_1^2 + \beta_3 X_2) \tag{11}$$

where $(\beta_1, \beta_2, \beta_3) = (0.4, 0.4, 1)$ and the function $\text{expit}(A) = 1/1 + \exp(-A)$. Because the selection model in (11) placed roughly half of the 2,000 schools in the experimental sample, we chose a random subset of 100 of these schools to be in the study sample ($Z = 1$). Of the sampled schools, half are randomized to treatment ($W=1$) and the other half to control ($W=0$). The potential outcomes under each treatment condition are given by:

$$Y(1) = \gamma_1 X_1 + \gamma_2 X_2 + \gamma_1 X_1^2 + \gamma_2 X_2^2 + 1$$

$$Y(0) = \gamma_1 X_1 + \gamma_2 X_2 \tag{12}$$

where $(\gamma_1, \gamma_2) = (0.1, 1)$. All potential outcomes were restricted to lie within $[Y^L, Y^U] = [-2, 2]$.

From (11) and (12), $X_1, X_2$ are the most predictive of sample selection and the treatment effect so propensity score stratification with these covariates is expected to yield the most precision gain. However, in studies where sampling ignorability is violated, the propensity score model may not include all the relevant covariates that affect both sample selection and the treatment effect. This raises the question of whether stratification would be effective in tightening bounds if alternative covariates (ones that do not necessarily moderate the treatment effect or affect sample selection) are used. To address this question, we considered five covariate combinations for the propensity score model: (a) $X_1, X_2$, (b) $X_3, X_4$, (c) $X_1, X_3$, (d) $X_2, X_4$ and (e) $X_1, X_2, X_3, X_4$. We also varied the correlation $\rho$: 0.25, 0.50, 0.70, to determine whether the effectiveness of propensity score stratification with alternative covariates depends on how strongly correlated the covariates are to $X_1, X_2$.

*Plausibility of MSS*

We conducted two simulation studies to assess the extent to which MSS contributes to the precision gain of bounds if plausible, and the implications for inference if MSS is not plausible. Under Simulation Study 1, MSS is plausible by restricting the potential outcomes in (12) to lie within [-1, 1] for schools in the population. Thus, a school that selects into the sample ($Z = 1$) has a larger expected outcome compared to a school in the population ($Z = 0$). Under Simulation Study 2, MSS is only weakly plausible where the potential outcomes for all schools (both in the sample and in the population) lie within the same [-2, 2] range. In Simulation 2, MSS only holds for less than 10% of the schools in the population[2].

*Redefining the Population of Inference*

In addition to the original population of inference $P$, we created three subpopulations based on the distribution of covariates $X = X_1, X_2, X_3, X_4$ in the sample. The subpopulations, $P_3, P_2, P_1$, are constructed by excluding population schools whose $X$ values lied outside three, two, and one standard deviations of the average $X$ values in the sample, respectively. For example, all population schools in $P_3$ have covariate values $X$ that lie within three standard deviations of the average covariate values in the sample. Note that the methods used here to define the subpopulations is one of several approaches and other methods of redefining the population can also potentially improve the precision of the bounds. For example, Tipton et al. (2017) constructed subpopulations based on specific covariates by excluding charter schools and schools with over 95% male students, among other variables. The goal is to create a target population of inference to which the results from the study sample is the most generalizable.

An important question is whether the redefined population still constitutes a relevant population for a given policy of interest. We offer two points of consideration: the size of the redefined population and the inclusion/exclusion of specific schools. In many generalization studies in education, the study sample is typically 5 – 10% of the population of inference. When the sample comprises a larger proportion of the redefined population, say half, but only 5% of the original population, it is possible that the redefined population may be compositionally different from the original population of inference. In this case, the researcher should compare the distributions of the covariates between the redefined population and the original population to determine the extent to which individuals in the populations are different. Second, if a policy of interest is focused on a specific group of students, the redefined population should include this group. For example, if the policy question pertains to both public and charter schools, but the redefined population excludes charter schools, this subpopulation may not be relevant for the research question of interest. These examples highlight the point that while redefinition may improve generalizations, it is also important to consider the balance between strengthened generalizations and the relevance of the redefined population to the policy questions of interest.

---

[2] The proportion of 10% was not controlled for in the simulation. This proportion was observed by counting the number of population schools for which MSS was plausible after the range of potential outcomes was set to [-2, 2].

**Additional Simulation Parameters**

*Sensitivity Parameter*

Our discussion of the bounding framework is situated around studies in which sampling ignorability does not hold. To incorporate violations to sampling ignorability, we apply a similar method to Coppock et al. (2017). We included a sensitivity parameter δ that represents the proportion of schools in the population for which sampling ignorability does not hold. The potential outcomes $Y(1), Y(0)$ are generated from a mixture model in which the population $P$ consists of schools that satisfy and violate sampling ignorability. Specifically, for a given fraction $\delta \in [0,1]$ of schools in $P$, the potential outcomes $Y(1), Y(0)$ are given by:

$$Y(1) = \gamma_1 X_5 + \gamma_2 X_6 + \gamma_1 X_5^2 + \gamma_2 X_6^2 + 1$$
$$Y(0) = \gamma_1 X_5 + \gamma_2 X_6 \qquad (13)$$

where $(\gamma_1, \gamma_2) = (0.1, 1)$ are the same coefficients as in (12) and the potential outcomes for the remaining $1 - \delta$ fraction of schools are given by (12). The potential outcomes in (13) were restricted to [-1, 1] for Simulation Study 1 and to [-2, 2] in Simulation Study 2 to be consistent with the MSS plausibility conditions. The continuous covariates $X_5, X_6$ were generated from a *t* distribution centered at zero with degrees of freedom *df* = 9 and 3, respectively. Under this framework, sampling ignorability is violated by omitting these variables from the propensity score models. The parameter $\delta$ ranged from 0 to 1 in increments of 0.20, where larger values of $\delta$ imply larger proportions of population schools where sampling ignorability does not hold.

*Alignment Scenarios*

Kern et al. (2016) considered the impact of covariate concordance in the selection and outcome model on the extent of bias reduction in propensity score-based estimators. The authors found that certain estimators performed better (in terms of bias) when there was positive concordance or "alignment" rather than negative alignment among the covariates. As an extension to this work, we include positive and negative alignment scenarios for both simulation studies to evaluate whether covariate alignment affects the precision of bounds, particularly under stratification. In the positive alignment scenario, the coefficients for $X_1, X_2$ were set to $(\beta_1, \beta_2, \beta_3) = (0.4, 0.4, 1)$ in the selection model in (11) and to $(\gamma_1, \gamma_2) = (0.1, 1)$ in the outcome model in (12). Thus, $X_2$ has a stronger impact in both the selection and outcome model. In the negatively aligned scenario, the coefficients were set to $(\beta_1, \beta_2, \beta_3) = (1, 0.5, 0.4)$ and $(\gamma_1, \gamma_2) = (0.1, 1)$ so that $X_2$ plays a weaker role in the sample selection model, but a stronger role in the outcome model.

**Simulation Results**

In this section, we present the results for Simulation Study 1 and 2 based on 100 replications for each combination of simulation parameters.[3] As a comparison, we also provide the SATE, the standard errors (SE), 95% confidence intervals, and average bias. We focus our discussion around two specific outcomes: (a) the widths of the worst-case and MSS bounds under all methods and (b) the coverage rates of the bounds and confidence intervals. We focus on bound width in much of our discussion to evaluate the effectiveness of different methods on improving precision in the bounding framework. We discuss coverage rates to compare the performance of the bounds and confidence intervals in situations where sampling ignorability does not hold.

---

[3] We calculated the Monte Carlo standard error (MCSE) of the estimated bounds and point estimates and determined the appropriate number of replications based on an error tolerance of 0.05. The average MCSE for each estimate was less than 0.05 with 100 replications.

*Precision Gain Under MSS and Redefinition of the Population*

Table 2 provides the bounds for Simulation Study 1 (MSS is plausible) and Simulation Study 2 (MSS is weakly plausible). The table is organized into four sets of rows, each corresponding to a different population of inference. We include the PATE for each value of $\delta$ for reference. Under $P$, the worst-case bounds range from about -3.0 to 3.0 in Simulation 1 and from about -3.6 to 3.7 in Simulation 2 across all values of $\delta$. This implies that the PATE ranges from about 3 standard deviations below zero to 3 standard deviations above zero in both studies. The worst-case bounds are wide because these bounds use the lower and upper end of the potential outcomes range. Under MSS, the bounds shrink to about [-3.0, 0.80] in both studies and are 40% narrower in width on average compared to the worst-case bounds across all values of δ. If MSS is plausible, this assumption contributes a large amount of identifying power. When δ increases, the estimated worst-case and MSS bounds are the same overall. This is because the bounds are mainly based on the sample, which is not affected by changing values of δ.

Surprisingly, when $P$ is redefined to $P_3$, the average bound width did not change much. In some cases, the bounds are wider by about 0.2% under $P_3$. Upon closer inspection, we found that the exclusion criteria used to construct $P_3$ excluded less than 3% of schools, so that this subpopulation did not constitute a significant redefinition of the population. As a result, most of the bounds under $P_3$ did not change much, in either simulation study. We see more of a reduction in bound width under $P_2$ and the largest reduction in $P_1$ where the bounds are narrower by 20% on average compared to $P$. Notably, the MSS bound under $P_1$ is narrower by about 50% compared to the worst-case bound in $P$, across all values of $\delta$ in both simulation studies. This illustrates that both MSS and redefining the population contribute to precision gains, but collectively, these gains can be even larger.

TABLE 2

*SATE Point Estimates*

Table 3 provides the SATE estimates, which are largely similar in both simulation studies. The PATE is larger on average in Simulation Study 2 because the range of potential outcomes is wider. As in Table 2, the SATE does not change much across all values of $\delta$ since the sample remains unchanged. Notably, the 95% confidence intervals for the PATE are narrower compared to the bounds and they are consistent with a positive PATE. Under $P$, as $\delta$ increases, the average bias of the SATE increases in both simulations. This is as expected because larger values of $\delta$ imply that larger proportions of population schools have different distributions of potential outcomes. When $P$ is redefined, the SATE remains unchanged, but the PATE changes as population schools are excluded. Importantly, the average bias is non-zero across all conditions.

TABLE 3

*Coverage Rates*

Figure 2 provides the coverage rates across all values of $\delta$ for the worst-case and MSS bounds and the 95% confidence intervals. For the sake of parsimony, we only provide the coverage rates for $P$ and describe the trends in the subpopulations. Under Simulation 1, the worst-case and MSS bounds have nearly 100% coverage across all values of $\delta$, but the coverage for the confidence intervals decreases with increasing $\delta$. The trend for the MSS bound is consistent with the framework of Simulation 1 where MSS is plausible. Under Simulation 2, the MSS bounds have poor coverage at less than 25% across all values of $\delta$. As a result, imposing this assumption when it is only weakly plausible is inappropriate. The coverage rate for the confidence intervals is higher than the MSS bounds in Simulation Study 2, but interestingly, it is still not 100%, even when $\delta = 0$. Thus, even if MSS is weakly plausible, the coverage rate of the confidence intervals may still be less than ideal.

The coverage rates are similar in the subpopulations, but there are two important differences. First, in Simulation 1, the coverage rate for the confidence intervals increases with larger values of $\delta$ in all subpopulations, but they are still below 60%. This is because under Simulation 1, the PATE decreases with increasing $\delta$, which improves the coverage rates of the confidence intervals. Second, for Simulation 2, the coverage rate of the MSS bounds is above 60% for small values of $\delta$ under $P_1$, but quickly drops below 25% as $\delta$ increases beyond 0.50. This is likely due to the combined effects of excluding population schools, which affects the PATE, and the weak plausibility of the MSS assumption in Simulation 2.

FIGURE 2

*Precision Gain Under Stratification*

Tables 4 and 5 provide the precision gain, measured by the percent reduction in bound width, under propensity score stratification. Because the trends were similar in both alignment scenarios, we focus on the positive alignment case and provide results under negative alignment in the Appendix (Tables i and ii). The first and second panel in the tables compare the worst-case to worst-case and the MSS to MSS bounds, before and after stratification, respectively. The third panel compares the worst-case bounds without stratification to the MSS bounds with stratification to assess the extent to which both MSS and stratification narrow bounds. Within each panel, the results are organized by the five covariate combinations.

As expected, under $P$ and $\delta = 0$, the largest precision gain is based on stratification with $(X_1, X_2)$ where the bound width was reduced by about 12% in both simulations. Surprisingly, in Simulation 1, the precision gains were similar across all covariate combinations, including $(X_3, X_4)$, in both the MSS and WC-MSS cases. In Simulation 2, the precision gain under $(X_3, X_4)$ is one of the smallest under MSS and WC-MSS, but the differences between covariate combinations are small. We suspect that since MSS contributes a large amount of identifying power alone, stratification does not contribute as much additional identifying power. Furthermore, stratification had the weakest effect in the MSS comparison where the average precision gain was less than 7% across all covariate combinations. This is in comparison to the 12% seen for the worst-case bounds. The largest gains are seen for WC-MSS where the bound widths are reduced by nearly 40%. Note that this is similar to the gains seen when MSS and redefining the population were combined.

After $(X_1, X_2)$, stratification with $(X_2, X_4)$ yielded the second largest precision gain in Simulation 1. Interestingly, stratification with $(X_2, X_4)$ yielded larger precision gains compared to $(X_1, X_2)$, which may be because $X_2$ plays a stronger role in both the selection and outcome models. On the other hand, stratification with $(X_1, X_3)$ yielded one of the lowest gains in both studies, possibly due to the omission of $X_2$ and the weaker role of $X_1$ in the selection and outcome models. As $\delta$ increases, the average gain decreases in both studies since the potential outcomes depend more on $X_5, X_6$. When $P$ is redefined, the average precision gain increases overall and the gains are comparatively larger under Simulation 2.

TABLES 4, 5

*Stratification and $\rho$*

Tables 4 and 5 only include the results for $\rho = 0.25$ since the trends for the other correlations were similar. For comparison, we include the results for $\rho = 0.70$ in the Appendix (Tables iii and iv). The purpose of varying the correlation is to assess whether stratification with alternative covariates improves precision and if so, whether the gains are affected by the magnitude of the correlation. From the tables for $\rho = 0.25$ and $\rho = 0.070$, the results are mixed, but the differences in correlation appears to have the largest effect on stratification with $(X_3, X_4)$. In both simulations, when $\rho$ increases from 0.25 to 0.70, the average precision gain based on $(X_3, X_4)$ also increases, but only by about 8% for the worst-case bounds under $\delta = 0$. When $\delta$ increases, the results are mixed where in some cases, the precision gain is the same for both $\rho = 0.25$ and $\rho = 0.70$. This is also seen under stratification with other covariate combinations. This suggests that increasing the correlation alone does not necessarily imply a larger precision gain and if it does, the gains are modest at best.

*Stratification and Point Estimates*

Tables 6 and 7 provide the SATE estimates under stratification. Overall, the trends are similar to the unstratified case, but there are two important differences. First, Table 6 shows that stratification *increased* the average bias in Simulation 1 across most of the covariate combinations. The exception is with $(X_3, X_4)$, where the bias is nearly the same magnitude as in the unstratified case. This is likely due to the differences in distribution of potential outcomes where schools in the population have outcomes that lie within a narrower range. As a result, stratification did not reduce the bias since the distribution of potential outcomes was still different within the strata. Second, stratification was more effective in reducing bias in Simulation 2, particularly under $(X_1, X_2)$ and $(X_2, X_4)$, and this result is consistent with the trends in Table 5 where stratifying by these covariates was associated with the largest reduction in bound width. Because the distribution of potential outcomes lied within the same range in this study, stratification was more effective with bias reduction, particularly when $\delta = 0$.

TABLES 6, 7

*Coverage Rates Under Stratification*

Figures 3 and 4 show the coverage rates under stratification with the original population $P$. The trends are similar overall to the unstratified case, but there are two exceptions. First, in Simulation 1, there is a steeper drop in coverage for the confidence intervals when stratification with $(X_3, X_4)$ is used. Similarly, in Simulation 2, the coverage rate of the MSS bounds is still below 50%, and it is lowest under stratification with $(X_3, X_4)$. Second, Figure 4 shows that the coverage rates for the confidence intervals are higher in the stratified case compared to the unstratified case across all covariate combinations. This is consistent with the results from Table 7 where stratification contributed to bias reduction. However, the coverage rate for some covariate combinations is still well below 100%.

FIGURES 3, 4

**Discussion**

The simulation results suggest several important implications. First, if alternative assumptions like MSS are plausible, these assumptions potentially contribute a large amount of identifying power in the bounding framework. In our study, MSS bounds were 40% narrower on average compared to the worst-case bounds. Second, combining MSS with redefining the population or with stratification can narrow the worst-case bounds further, to nearly 50% of their original bound width. Both methods serve to leverage the homogeneity among schools to increase the precision in the bound estimates. Third, the precision gain associated with propensity score stratification depends on the covariates in the model and the gains are largest when the covariates are predictive of both sample selection and the outcome. However, our results suggest that stratification may not contribute as much additional gain when assumptions like MSS are invoked. This is likely because most of the precision gain is associated with the assumption itself. Additionally, when alternative covariates are used, the effectiveness of stratification increases when the correlation among covariates increases, but the gains are small. In our simulations, increasing the correlation from 0.25 to 0.70 only increased the precision gain by about 8% in both studies. Lastly, assumptions like MSS can have a large impact on bounds, but this impact depends on the plausibility of the assumption. The coverage rates illustrate that the MSS bounds have nearly 100% coverage when plausible, but these rates quickly fall below 50% when the assumption is weakly plausible. Additionally, when sampling ignorability is violated, the coverage rates for the confidence intervals for the SATE may still be less than 100%. We provide the coverage rates to highlight the importance of assessing the implications for inference when core assumptions for generalization do not necessarily hold.

**Application to SimCalc**

We return to SimCalc and re-analyze the study using the bounding framework illustrated in this article. The population, *P*, consists of the *N*=1,713 non-charter middle schools in Texas during the 2008 – 2009 academic year with the probability of selection estimated as $\Pr(Z = 1) = 92/1713 \approx 0.054$. The outcomes, *Y*, are the aggregate student gain scores, standardized by the between school variance. For the purpose of illustration, we set the lower and upper bounds, $[Y^L, Y^U]$ to $[-2, 3]$ using the observed range of the gain scores from the sample. The PATE is defined as the difference in average standardized gain scores among schools that used SimCalc compared to schools under a "business as usual" condition. For the stratified bounds, we used the

26 covariates in Table 1 to estimate the propensity scores and stratify the population into five equally sized strata.

To estimate the precision of the bounds, we provide bootstrapped confidence intervals. This approach was considered in Hardle (1990) and Manski et al. (1992), and more technical approaches have been discussed in Canay and Shaikh (2016). The confidence intervals are estimated as follows: (1) take repeated samples of size *n* and estimate the bounds under each framework (worst-case, MSS, and stratified versions of each case) for each bootstrapped sample; (2) repeat the first step 1000 times, and then take the 0.05 quantile of the bootstrapped distribution of the lower bound and the 0.95 quantile of the bootstrapped distribution of the upper bound. The bootstrapped bounds have 95% confidence for containing the *true* PATE if the estimation uncertainty is on the same order or smaller than the true bound width. Note that although we use the 0.05 and 0.95 quantiles of the bootstrapped distributions, the resulting bounds still achieve 95% confidence for containing the true parameter. This is because any error from coverage is one-sided. If the true PATE was close to the lower end of the bound, we would focus on coverage at the lower end and the same would be true if the PATE was closer to the higher end of the range. Because of the focus on one side of the range, the level of significance α does not need to be split across the intervals. In the results, the bootstrapped bounds are denoted by $[LB_{0.05}, UB_{0.95}]$.

Table 8 provides the bounds for the PATE under the worst-case and MSS frameworks for four different populations. We estimate the bounds for the original population *P*, and three subpopulations $P_{sub}$, $P_3$ and $P_2$ based on several redefinitions of *P*. The subpopulation $P_{sub}$ is identical to the population defined in Tipton (2013) and excludes population schools whose estimated propensity scores lie outside the range of the propensity scores for the SimCalc study schools. This reduced the original number of schools from 1,713 to $N_{sub} = 1,581$. Following the approach in the simulation study, the subpopulations $P_3$ and $P_2$ were constructed by excluding population schools whose covariate distributions (based on the 26 covariates) lied outside three and two standard deviations of the sample means. We omitted $P_1$ as the size of this subpopulation was smaller than the SimCalc study. For all subpopulations, we refit the sampling propensity score model after omitting schools that do not meet the inclusion criteria and subset the original population $P$ based on the refitted propensity scores $s(X)$. Note that the estimand of interest is different for each subpopulation. In particular, the estimand is the expected treatment impact that is specific to the types of schools included in each subpopulation, which may be different than the expected treatment impact for the overall population that includes all schools.

From Table 8, the unweighted bounds (without stratification) on the PATE under $P$ have an average width of 9.8 under the worst-case framework and the bounds shrink by 36% under MSS. Both bounds are wide and consistent with an estimated PATE of zero, where the average difference in student gain scores ranges from 4.840 standard deviations below to 4.995 and 1.438 standard deviations above zero for the worst-case and MSS frameworks, respectively. Both sets of bounds narrow as the population is redefined, achieving the tightest width under $P_2$ when all the included population schools have covariates that are within 2 standard deviations of the sample. Although all bounds under each subpopulation are consistent with an insignificant PATE, both MSS and redefining the population contribute identifying power to narrow the range of values.

TABLE 8

*Bounds for SimCalc by Propensity Score Strata*

To tighten the bounds further, we stratified the population schools using estimated propensity scores based on the covariates in Table 1. The stratum-specific bounds are provided for each population and the final bounds are listed as "Stratified Bounds" in Table 8. Across each population, stratification narrowed the worst-case and MSS bounds by as much as 38%, which is similar to the amount of precision gain that MSS contributed in the unweighted bounds. This reduction is much larger than the precision gains seen in the simulation studies, where stratification was associated with only a 12% reduction in bound width for the worst-case bounds and with less than 10% for the MSS bounds. For the worst-case bounds under $P$, the results illustrate that combining MSS and stratification can narrow the bound width by as much as 50%, which is larger than the 40% gains seen in the simulation studies. Although all the stratified bounds are wide, stratification provides a potentially useful approach to improve precision, particularly when combined with additional assumptions on the observed data. Finally, the confidence intervals for the unweighted and stratified bounds are given in the fifth and eighth column of Table 8 for the worst-case and MSS bounds, respectively. Incorporating estimation uncertainly only slightly expands the bounds for the worst-case and MSS framework so that the overall trends seen above were replicated in the bootstrapped bounds.

**Discussion**

Policymakers often use results from well-designed evaluations to inform decisions, but little attention has been paid to the underlying assumptions in the statistical analysis. The fact that core assumptions may be violated in practice has implications for the validity of inferences, particularly when the study results may be applicable to only a subset of the population. Sampling ignorability is an important assumption in generalization studies with nonrandom samples, but this assumption is violated whenever important covariates are omitted or when there exist population schools that share no overlap with the sample. When key assumptions are not met, the researcher must evaluate the advantages and tradeoffs of different methods to address violations in the assumptions.

In this article, we introduce and discuss the advantages and limitations of bounding, specifically for generalization studies where sampling ignorability is not necessarily satisfied. While bounding offers a different perspective to estimation in generalization, one important limitation to bounding frameworks is that even with methods such as stratification and redefining the population, the bounds can still be uninformatively wide. This is not unique to generalization but the bounds for generalization studies are often wide because the probabilities of sample selection are typically small. The bounding framework illustrated in this article does not substitute for the point estimates under sampling ignorability or the existing approaches based on sensitivity analysis and redefinitions of the population. However, we believe it is important for researchers to consider the tradeoffs of different methods, including bounding, when the validity of inferences relies on key assumptions that may not necessarily hold in practice.

In general, without probability sampling, estimation in generalization studies is limited and as a result, assumptions must be made. An important question is how researchers can evaluate the plausibility of these assumptions in practice. The challenge is that many of the assumptions, such as sampling ignorability and MSS for bounding, are untestable and their plausibility can only be suggested, not validated, by empirical evidence. The arguments we presented for MSS' plausibility relied on knowledge of the context of the study; namely, on knowledge of the relationships between the schools and the research centers. To evaluate the plausibility of any assumption, we believe that attention to context is important as different assumptions may be plausible in different situations. Additionally, context is crucial when interpreting the treatment effects for a given study (Lemons et al., 2014). Future research should review the types of assumptions made in current studies and evaluate the extent to which prior evidence or empirical evidence from different fields support the plausibility of the assumptions.